\documentstyle[11pt,draft]{article}
\newcommand{\be}{\begin{eqnarray}}
\newcommand{\ee}{\end{eqnarray}}
\newcommand{\bra}[1]{\mbox{$\langle\, #1 \mid$}}

\newcommand{\ket}[1]{\mbox{$\mid #1\,\rangle$}}

\newcommand{\pro}[2]{\mbox{$\langle\, #1 \mid #2\,\rangle$}}
\newcommand{\expec}[1]{\mbox{$\langle\, #1\,\rangle$}}
\newcommand{\expecl}[1]{\mbox{$\left\langle\, 
            \strut\displaystyle{#1}\,\right\rangle$}}

\renewcommand{\a}{\hat a}
\newcommand{\ac}{\hat a^\dagger}
\renewcommand{\b}{\hat b}
\newcommand{\bc}{\hat b^\dagger}
\title{Adiabatic Invariant Treatment of a Collapsing Sphere 
of Quantized Dust}
\author{Roberto Casadio\thanks{e--mail: casadio@bo.infn.it}, 
\ Fabio Finelli\thanks{e--mail: finellif@bo.infn.it} 
\ and Giovanni Venturi\thanks{e--mail: armitage@bo.infn.it}\\
 \\
{\em Dipartimento di Fisica, Universit\`a di
Bologna} \\
{\em and} \\
{\em Istituto Nazionale di Fisica Nucleare, 
Sezione di Bologna, Italy}}
\begin{document}
\begin{titlepage}
\pagestyle{empty}
\maketitle
\begin{abstract}
The semiclassical collapse of a sphere of quantized dust is studied. 
A Born-Oppenheimer decomposition is performed for the wave 
function of the system and the semiclassical limit is considered 
for the gravitational part.
The method of adiabatic invariants for time dependent Hamiltonians is 
then employed to find (approximate) solutions to the quantum dust
equations of motions.
This allows us to obtain corrections to the adiabatic approximation 
of the dust states associated with the time evolution of the metric. 
The diverse non-adiabatic corrections are generally associated with 
particle (dust) creation and related fluctuations. 
The back-reaction due to the dominant contribution to particle 
creation is estimated and seen to {\sl slow-down} the collapse.  
\end{abstract}
\end{titlepage}
%
\pagestyle{plain}
\raggedbottom
\setcounter{page}{1}
\section{Introduction}
The canonical quantization of general relativistic isotropic 
systems carried out in suitably chosen variables leads
to the whole dynamics being determined by the Hamiltonian
constraint of the Arnowitt, Deser and Misner (ADM) construction.
Such an approach is particularly useful if one wishes to study
the semiclassical regime of a system of self gravitating 
matter and has been applied to the collapse of a sphere of classical 
dust with the associated scalar field being related to time 
\cite{lund}.
In a previous paper \cite{cv}, two of the present authors 
again applied the ADM formalism to a model describing the collapse 
of a sphere of homogeneous dust \cite{oppenheimer} leading to a black 
hole and examined how Hawking radiation arises within such an approach.
In particular the Born-Oppenheimer (or adiabatic) approximation 
for the coupled matter-gravity system was consistently implemented
according to the original formulation of Refs.~\cite{brout,bv}
(for alternative views see also \cite{kiefer,kim,datta} and for
their comparison see \cite{bertoni})
by assuming the gravitational degree of freedom evolved 
{\sl slowly} with respect to the matter degree of freedom.
In this note we shall study corrections to this approximation.
\par
The classical collapse of an isotropic homogeneous sphere of dust 
\cite{oppenheimer}
has been studied extensively and is treated in numerous text books
({\em e.g.}, \cite{stephani}).
A particularly interesting aspect of such a collapse is that the interior 
of the sphere is a three dimensional space of constant curvature whose
radius depends on time (in the language of cosmological models it is a
section of a Friedmann universe).
Further, it has recently been shown that boundary effects are 
in general absent for classical fluids with a step-function discontinuity 
of the kind associated with such models \cite{hk} (see also \cite{os} 
for a more specific treatment of the sphere of dust).
The treatment of a scalar field in a cosmological context is
known and for a particular regime one has that the homogeneous mode of 
the free scalar field is related to dust, that is a fluid with constant 
density and zero pressure \cite{madsen,parker}.
One may then naturally ask to what extent is this possible for the
spherically symmetric collapse of a massive scalar field, that is
what is the effect of the boundary of the sphere.
Obviously if the radius of the sphere is infinite one must
reproduce the cosmological models.
Therefore, whatever the difference is between the two cases, 
it must vanish as the radius of the sphere tends to infinity.
Since a free massive scalar field is localized within its Compton 
wavelength and just ``feels'' objects at such a distance, 
the boundary will only affect fields a Compton wavelength away and 
its net effect on the sphere of matter will be proportional to the 
ratio of the Compton wavelength of the scalar field to the radius 
of the sphere.
The former ratio is expected to be small for sufficiently large spheres
(thus 3-curvature is relatively small),
a condition which has been previously noted and related to a breakdown
of the adiabatic approximation for dust \cite{cv} or to the validity
of the semiclassical (WKB) approximation \cite{moss}.
\par
Let us illustrate the above consideration by a simple model.
In the comoving frame the dust particles are at rest and the adiabatic
approximation implies that the radius of the sphere can be kept 
approximately fixed \cite{cv} on solving the matter equation of motion.
Let us then consider a static spherically symmetric space-time metric  
\be
ds^2=-d\tau^2+dr^2+r^2\,d\Omega^2
\ ,
\ee
which of course corresponds to a flat 3-space and also agrees
with the Schwarzschild vacuum far from the event horizon.
The action for a spherically symmetric scalar field $\Phi(r,\tau)$
with inverse Compton wavelength $\mu$ will be given by
\be
S_\Phi= 4\,\pi\,\int d\tau\,\int r^2\,dr\,{1\over 2}\,
\left[(\dot\Phi)^2-(\Phi')^2-\mu^2\,\Phi^2-2\,V\,\Phi^2\right]
\ ,
\label{toy}
\ee
where $\ '\equiv d/dr$, $\dot{\ } \equiv d/d\tau$ and we have introduced
a potential $V$ which is responsible for the spherical confinement of the
scalar field and is zero inside the sphere ($r<r_0$).
The equation of motion for $\Phi$ is given by
\be
\ddot\Phi-\Phi''-{2\over r}\,\Phi'+\mu^2\,\Phi+2\,V\,\Phi=0
\ .
\ee
Since we seek an analogy for dust, we want a solution that is homogeneous
for $r<r_0$ and zero for $r\gg r_0$ (no matter outside the sphere).
Such a solution is given by
\be
\Phi=\phi(\tau)\,{f(r)\over r}
\ ,
\ee
with $\ddot\phi+\mu^2\,\phi=0$, since our solution is stationary, and
\be
f=\left\{\begin{array}{ll}
r & \ \ \ r<r_0 
\\
\\
r_0\,e^{-\mu\,(r-r_0)} & \ \ \ r>r_0
\ ,
\end{array}\right.
\label{f}
\ee
in which we have made the choice of $f/r$ dimensionless (so that 
$\phi$ has the same dimensions as $\Phi$) and any 
other dimensionless factor has been absorbed into $\phi$.
The corresponding potential is then for $r>r_0$
\be
V={\mu^2\over 2}
\ ,
\ee
plus a Dirac delta singularity at $r=r_0$ which ensures the 
continuity of the derivatives of $f$.
One may second quantize $\Phi$, by introducing creation ($\ac$) and 
destruction ($\a$) operators for the above solution, obtaining 
($m_\phi\equiv\hbar\,\mu$ is the mass of a scalar quantum)
\be
\hat\Phi\sim\sqrt{{m_\phi\,\mu\over 2}}\,\left[
\ac\,e^{-i\,\mu\,\tau}+\a\,e^{i\,\mu\,\tau}\right]
\,{f\over r}
\ ,
\ee
and subsequently evaluate quantities of physical interest for 
any state of dust quanta.
One finds for the energy density \cite{madsen,parker}
\be
\tilde\rho&=&{1\over 2}\,
\expec{(\dot{\hat\Phi})^2+({\hat\Phi'})^2+\mu^2\,\hat\Phi^2
+2\,V\,\hat\Phi^2}
\nonumber \\
&\sim&
m_\phi\,\mu\,\expec{\ac\,\a+{1\over 2}}\,
\left[(\mu^2+V)\,{f^2\over r^2}
+{1\over 2}\,{\left({f\over r}\right)'}^2\right] 
\nonumber \\
&\sim&
m_\phi\,\mu^3\,\expec{\ac\,\a+{1\over 2}}\,
\left\{\begin{array}{ll}
1 &\ \ \ r<r_0 
\\
\\
2\,{r_0^2\over r^2}\,e^{-2\,\mu\,(r-r_0)}\, 
\left(1+{1\over 2\,\mu\,r}+{1\over 4\,\mu^2\,r^2}\right)
&\ \ \ r>r_0
\ ,
\end{array}\right.
\ee
which corresponds to a constant density inside the sphere and an 
exponentially decreasing density with range $\sim 1/\mu$ outside, 
as desired.
Lastly one may evaluate the pressure $P$ \cite{madsen,parker} 
obtaining 
\be
P&=&{1\over 2}\,
\expec{(\dot{\hat\Phi})^2-{1\over 3}\,({\hat\Phi'})^2-\mu^2\,\hat\Phi^2
-2\,V\,\hat\Phi^2}
\nonumber \\
&\sim&
-m_\phi\,\mu^3\,\expec{\ac\,\a+{1\over 2}}\,
\left\{\begin{array}{ll}
0 &\ \ \ r<r_0 
\\
\\
{2\over 3}\,{r_0^2\over r^2}\,e^{-2\,\mu\,(r-r_0)}\,
\left(1+{1\over2\,\mu\,r}+{1\over4\,\,\mu^2\,r^2}\right)
&\ \ \ r>r_0
\ ,
\end{array}\right.
\ee
which is zero inside the sphere and its magnitude decreases exponentially 
outside.
\par
We may at this point, as in cosmological models, obtain an effective
action for $\phi(\tau)$ by substituting the particular spatial function 
$f(r)$, Eq.~(\ref{f}), in the original Eq.~(\ref{toy}) and integrating over 
the spatial coordinate $r$:
\be
S_\Phi&\sim&4\,\pi\,\int d\tau\,\int r^2\,dr\,{1\over 2}\,
\left[\left(\dot\phi^2-\mu^2\,\phi^2\right)\,{f^2\over r^2}
-\phi^2\,{\left({f\over r}\right)'}^2
-2\,V\,\phi^2\,{f^2\over r^2}\right]
\nonumber \\
&\sim&
{4\,\pi\,r_0^3\over 3}\,\int d\tau\,\left[{1\over 2}\,
\left(\dot\phi^2-\mu^2\,\phi^2\right)
+{\cal O}\left({1\over\mu\,r_0}\right)\right]
\equiv S_\phi+{\cal O}\left({1\over\mu\,r_0}\right)
\ .
\label{eff}
\ee
Thus, unless one has $r_0\sim1/\mu$, the edge effects are negligible
as we heuristically indicated before.
Of course in the above simple model we have not considered all possible 
modes for the scalar field, but only the mode corresponding to dust and we 
have seen that for such a mode upon quantization one obtains, for any 
state, pressureless dust inside the sphere \cite{cv}.
\par
In section~\ref{review} we briefly review the effective action used
to describe the collapsing sphere of dust along the lines suggested 
above (which will allow us to maintain the Robertson-Walker form of 
the metric) and the results obtained in Ref.~\cite{cv},
in order to prepare the ground for the study of the back-reaction on 
the geometry induced by quantum evolution (collapse) of the dust.
This will require relaxing the adiabatic approximation and is the aim 
of the present note.
\par
In section~\ref{collapse} the method of adiabatic invariants 
\cite{lewis-ries,gao} is illustrated and is applied to solve 
the matter Schr\"odinger equation with a time dependent Hamiltonian 
obtained through the Born-Oppenheimer decomposition of the 
matter-gravity wave function and in the semiclassical limit for 
gravitation. 
After the identification of the parameter describing the adiabatic
limit and the associated states, coherent states for the matter wave 
function having correct classical limits are constructed and
used to determine the expectation value of the diverse quantities of 
physical interest. 
Since the collapse is adiabatic at some initial time, from the 
expressions obtained we estimate and compare the subsequent 
deviations from the adiabatic approximation and see the effect 
of back-reaction on gravitation.
Lastly in section~\ref{concl} our results are summarized and 
discussed.
\par
We use units for which $c=1$, $\kappa\equiv 8\,\pi\,G$,
the Planck length is then $\ell_p\equiv\sqrt{\kappa\,\hbar}$
and the Planck mass is $m_p=\hbar/\ell_p$.
\setcounter{equation}{0}
\section{Semiclassical collapse of a sphere of dust}
\label{review}
\par
As mentioned in the Introduction, in order to describe the evolution of 
a collapsing sphere of homogeneous dust in vacuum it is convenient to 
consider a time-dependent scalar field $\phi$ confined inside
a spherical portion of a Robertson-Walker space-time 
(the interior of the sphere) with line element
\be
ds^2=K^2(\eta)\,\left[-d\eta^2+{d\rho^2\over 1-\epsilon\,\rho^2}
+\rho^2\,d\Omega^2\right]
\ ,
\ \ \ \ \ 0\le\rho\le\rho_0
\ ,
\label{g_hom}
\ee
where $\epsilon=0,\pm 1$. 
The radius of the sphere then follows from the matching condition 
with the external Schwarzschild metric and is given by $r_0=\rho_0\,K$ 
\cite{cv,stephani}.
The corresponding total effective action (gravity plus matter) is given 
by \cite{n_cv}
\be
S={1\over2}\,\int d\tau\,\left[
-{1\over\kappa}\,\left(K\,{\dot K}^2-\epsilon\,K\right)
+K^3\,\left({\dot\phi}^2-{\mu}^2\,{\phi}^2\right)
\right]
\ ,
\label{s1}
\ee
where $d\tau=K\,d\eta$ is the proper time of an observer comoving with 
the dust.
\par
The Hamiltonian obtained from the action in Eq.~(\ref{s1}) is
\be
H=-{1\over2}\left(\kappa\,{{\pi_{_K}}^2\over K}
+{\epsilon\over\kappa}\,K\right)
+{1\over2}\,\left({{\pi_{\phi}}^2\over K^3}
+{\mu}^2\,{\phi}^2\,K^3\right)
\equiv H_{_G}+H_{_M}
\ ,
\ee
where $\pi_{_K}=-\,K\,\dot K/\kappa$ and $\pi_{\phi}=K^3\,\dot\phi$.
Canonical quantization leads to the Wheeler-DeWitt equation,
\be
{1\over 2}\,\left[{\kappa\,\hbar^2}\,
\frac{\partial^2}{\partial K^2} {1\over K}
-{\epsilon\over\kappa}\,K
-{\hbar^2\over K^3}\,
{\partial^2\over\partial\phi^2}
+\mu^2\,\phi^2\,K^3
\right]\,\Psi(K,\phi)=0
\ ,
\label{wdw}
\ee
where we have chosen a suitable {\em operator ordering} in the 
gravitational kinetic term \cite{cv}. One now expresses $\Psi$ in the 
factorized form $\Psi(K,\phi)=K\,\psi(K)\,\chi(\phi,K)$
which, after multiplying on the l.h.s. of Eq.~(\ref{wdw})
by $\chi^\ast$ and integrating over the matter degrees of freedom,
leads to an equation for the gravitational part \cite{cv},
\begin{eqnarray}
& &{1\over2}\left[
\left({\kappa\,\hbar^2}\,{\partial^2\over\partial K^2}
-{\epsilon\over\kappa}\,K^2\right)
+{1\over\pro{\tilde\chi}{\tilde\chi}}\,
\bra{\tilde\chi}\,
\left({{\hat\pi_{\phi}}^2\over K^2}
+{\mu}^2\,\phi^2\,K^4\right)
\,\ket{\tilde\chi}\right]\,\tilde\psi
\nonumber\\   
& &\equiv\left[\hat H_{_G}\,K+K\,\expec{\hat H_{_M}}
\right]\,\tilde\psi 
=\frac{\kappa\,\hbar^2}{2 \pro{\chi}{\chi}}
\langle \chi |
\frac{\stackrel{\leftarrow}{\partial}}{\partial K}
\left( 1 - \frac{\ket{\chi} \bra{\chi}}{\expec{\chi|\chi}} \right) 
\frac{\partial}{\partial K} | \chi \rangle 
\nonumber\\
& &=-\frac{\kappa\,\hbar^2}{2\,\pro{\tilde\chi}{\tilde\chi}}
\bra{\tilde\chi}\,\frac{\partial^2}{\partial K^2}\,\ket{\tilde\chi}
\,\tilde\psi
\equiv-{\kappa\,\hbar^2\over 2}\,\expecl{\frac{\partial^2}{\partial K^2}}
\,\tilde\psi
\ ,
\label{wdw_g}
\end{eqnarray}
where we have defined a scalar product
\be
\pro{\chi}{\chi}\equiv
\int d\phi\,{\chi}^\ast(\phi,K)\,
\chi(\phi,K)
\ ,
\ee
and we have set
\be
\begin{array}{lcr}
\psi=e^{-i\,\int^K A(K')\,dK'}\,\tilde\psi &
\ \ \ \ \ \ &
\chi=e^{+i\,\int^K A(K')\,dK'}\,\tilde\chi
\ ,
\end{array}
\ee
with
\be
A\equiv \frac{-i}{\pro{\chi}{\chi}}
\bra{\chi}\,\frac{\partial}{\partial K}\,\ket{\chi}
\equiv -i \, \expecl{\frac{\partial}{\partial K}}
\ .
\label{berry}
\ee
If we now multiply Eq.~(\ref{wdw_g}) by $\tilde\chi$ and subtract it
from Eq.~(\ref{wdw}) we obtain \cite{cv}
\begin{eqnarray}
& &\tilde\psi\,K
\,\left[\hat H_{_M}-\expec{\hat H_{_M}}\right]\,\tilde\chi
+{\kappa\,\hbar^2}\,
\left({\partial\tilde\psi\over\partial K}\right)\,
{\partial\tilde\chi\over\partial K}
={\kappa\,\hbar^2\over2}\,\tilde\psi\,\left[
\expecl{{\partial^2\over\partial K^2}}-
\frac{\partial^2}{\partial K^2}\right]\,\tilde\chi 
\nonumber \\ 
&&=-\frac{\hbar^2 \kappa}{2}\,\psi\,\left[ \left(
\frac{\partial^2}{\partial K^2} 
-\expecl{\frac{\partial^2}{\partial K^2}}\right) 
-2\,\expecl{\frac{\partial}{\partial K}}\left(
\frac{\partial}{\partial K} - \expecl{\frac{\partial}{\partial K}}
\right) \right]\chi
\ ,
\label{wdw_m}
\end{eqnarray}
which is the equation for the matter (scalar field) wave function.
\par
On neglecting the r.h.s. of Eq.~(\ref{wdw_g}) one may introduce time
\cite{banks,brout,bv} by writing a semiclassical (WKB) approximation 
for the wave function $\tilde\psi$:
\be
\tilde\psi \simeq 
\left({\partial S_{eff}\over\partial K}\right)^{-1/2}
\,e^{+{i\over\hbar}\,\strut\displaystyle S_{eff}}
\ ,
\label{wkb}
\ee
where $S_{eff}$ is the effective action satisfying
the Hamilton-Jacobi equation associated with the
l.h.s. of Eq.~(\ref{wdw_g}),
\be
S_{eff}=-{1\over2\,\kappa}\,\int d\tau\,\left[
K_c\,{\dot K_c}^2-\left(\epsilon\,K_c-2\,\kappa\,\expec{\hat H_{_M}}
\right)\right]
\ ,
\label{s_eff}
\ee
and $\expec{\hat H_{M}}$ is now evaluated for $K=K_c$
which is where $\tilde\psi$, Eq.~(\ref{wkb}), has support.
One may then define a (proper) time variable,
\begin{eqnarray}
\frac{\partial\tilde\psi}{\partial K}\, 
\frac{\partial}{\partial K} &\simeq& 
\left[\frac{i}{\hbar}\,\frac{\partial S_{eff}}{\partial K}
-\frac{1}{2}\,\left(\frac{\partial S_{eff}}{\partial K}\right)^{-1} 
\,\frac{\partial^2 S_{eff}}{\partial K^2}\right]\,\tilde\psi\,
\frac{\partial}{\partial K}
\nonumber \\
&\equiv&  
-\frac{i\,K}{\kappa\,\hbar}\,\tilde\psi\,\frac{\partial}{\partial \tau}
-\frac{1}{2}\,\left(\frac{\partial S_{eff}}{\partial K}\right)^{-1}
\,\frac{\partial^2 S_{eff}}{\partial K^2}\,\tilde\psi\,
\frac{\partial}{\partial K}
\ .
\label{eta}
\end{eqnarray}
Further if the r.h.s. of Eq.~(\ref{wdw_m}) and the second term 
in the r.h.s. of Eq.~(\ref{eta}) are negligible,
one obtains the Schr\"odinger equation,
\be
i\,\hbar\,{\partial\chi_s\over\partial\tau}
={1\over 2}\,\left[-{\hbar^2\over {K_c}^3}\,\partial^2_\phi
+\mu^2\,{K_c}^3\,\phi^2\right]\,\chi_s
=\hat H_{_M}\,\chi_s
\ ,
\label{schro}
\ee
where we have scaled the dynamical phase
\be
\chi_s\equiv\tilde\chi\,\exp\left\{-{i\over\hbar}\,\int^\tau
\expec{\hat H_{_M}(\tau')}\,d\tau'\right\} 
\ ,
\label{dynamic}
\ee
and omitted $\tilde\psi$ while setting $K=K_c$, which is where the 
semiclassical gravitational wavefunction has support. 
\par
In Ref.~\cite{cv} Eq.~(\ref{schro}) was solved by making 
the adiabatic approximation $\dot K_c/ K_c\ll 1$
and evaluating $\expec{\hat H_M}$ for such solutions.
Eq.~(\ref{s_eff}) then becomes the action
of the Oppenheimer-Snyder model \cite{oppenheimer},
\be
S_{cl}=-{1\over2\,\kappa}\,\int d\tau\,\left[
K_{cl}\,{\dot K_{cl}}^2-\left(\epsilon\,K_{cl}-2\,K_0\right)
\right]
\ ,
\label{s_cl}
\ee
with $K_0$ constant and $K_{cl}$ is one of the classical trajectories
\be
\begin{array}{lcr}
\left\{
\begin{array}{l}
K_{cl}=K_0\,\partial_\eta h_\epsilon(\eta)   \\
 \\
\tau=K_0\,h_\epsilon(\eta)
\end{array}
\right.
&\ \ \ \ \ \ \  &
h_\epsilon(\eta)=\left\{\begin{array}{ll}
\eta-\sin\eta &\ \ \ \ \epsilon=+1 \\
& \\
\eta^3/6      &\ \ \ \ \epsilon=0  \\
& \\
\sinh\eta-\eta&\ \ \ \ \epsilon=-1
\ .
\end{array}\right.
\end{array}
\label{K_cl}
\ee
The matching condition between the inner metric and the
external Schwarzschild metric then gives for the mass parameter 
of the sphere
\be
M/\kappa=K_0\,{\rho_0}^3/\kappa=\expec{\hat H_{_M}}\,{\rho_0}^3
=N\,m_\phi\,{\rho_0}^3
\ ,
\label{junction}
\ee
where $N=\expec{\hat H_M}/m_\phi$ is the (constant) number of dust 
particles.
\par
Let us emphasize that Eq.~(\ref{junction}) introduces the size
of the ball in the model.
If, as suggested in the introduction (as well as in Refs.~\cite{cv,moss}), 
boundary effects are negligible, the above is the only difference
with respect to cosmological models where there is no exterior metric.
Further the adiabatic approximation amounts to 
$\expec{\hat H_M}=K_0\,\rho_0^3$ being constant, that is the number 
of dust particles is constant, although the frequency $\omega_c
=K_{cl}\,\expec{\hat H_M}$ of the corresponding Schr\"odinger state 
in the conformal time $\eta$ is not \cite{cv} (of course 
in the proper time the frequency $=\expec{\hat H_M}$ is constant).
\par
In Ref.~\cite{cv} it was subsequently verified that all the 
approximations were consistent for the collapse of the sphere up to 
its horizon radius $r_{_H}=2\,M$ if the Compton wavelength of the
scalar field is much smaller than the Schwarzschild radius of the
sphere,
\be
2\,M \gg \frac{1}{\mu}
\ ,
\label{cond}
\ee
which is precisely the condition $\mu\,r_0>2\,\mu\,M\gg1$ that one needs 
in order to neglect edge effects \cite{moss} (see Eq.~(\ref{eff})).
\par
In the following section we shall obtain solutions to Eq.~(\ref{schro}) 
without making the adiabatic approximation, thus allowing for a change 
in the number of dust particles, and shall estimate the
corresponding corrected classical trajectories $K_c$ \cite{cita}.
\setcounter{equation}{0}
\section{Adiabatic invariants and quantized matter}
\label{collapse}
A suitable method for the study of time dependent quantum systems is that of 
adiabatic invariants.
In particular given a time dependent Hamiltonian $\hat H_M(\tau)$, 
a Hermitian operator 
$\hat I(\tau)$ is called an {\sl adiabatic invariant} if it satisfies
\cite{lewis-ries,gao}
\be
i\,\hbar\,{\partial \hat I(\tau)\over d\tau}+
[\hat I(\tau),\hat H_M(\tau)]=0
\ .
\ee
The general solutions to the Schr\"odinger equation,
\be
i\,\hbar\,{\partial\chi_s(\tau)\over\partial\tau}
=\hat H_M(\tau)\,\chi_s(\tau)
\ee
are then given by
\be
\ket{\chi(\tau)}_{Is}=\sum\limits_n C_n\,
e^{i\,\varphi_n(\tau)}\,\ket{n,\tau}_I
\ ,
\ee
where $\ket{n,\tau}_I$ is an eigenvector of $\hat I(\tau)$ with 
time-independent eigenvalue $\lambda_n$ and the $\{C_n\}$ are complex
coefficients.
The phase $\varphi_n$ is given by
\be
\varphi_n(\tau)={i\over\hbar}\,\int_{\tau_0}^\tau\ _I
\bra{n,\tau'}\,\hbar\,\partial_{\tau'}+i\,\hat H_M(\tau')\,\ket{n,\tau'}_I
\, d\tau'
\ ,
\ee
and is the sum of the geometrical phase whose associated connection 
is given in Eq.~(\ref{berry}) and the dynamical phase displayed in 
Eq.~(\ref{dynamic}).
\par
The Hamiltonian in Eq.~(\ref{schro}) corresponds to a harmonic
oscillator of fixed frequency $\mu$ and variable mass
${K_c}^3$.
In this case it is useful to introduce the following 
linear (non-hermitian) invariant
\be
\hat I_b(\tau)\equiv e^{i\,\Theta(\tau)}\,\b(\tau)
\ ,
\ee
where the phase $\Theta$ is given by
\be
\Theta(\tau)=\int_{\tau_{_0}}^\tau 
{d\tau'\over {K_c}^3(\tau')\,x^2(\tau')}
\ ,
\label{theta}
\ee
and the operator $\b$ by
\be
\b(\tau)\equiv \frac{1}{\sqrt{2\,\hbar}}\,\left[{\hat \phi\over x}+
i\,\left(x\,\hat \pi_\phi-\dot x\,{K_c}^3\,\hat \phi\right)\right]
\ .
\ee
The function $x=x(\tau)$ is to be determined as a solution of the 
nonlinear equation
\be
\ddot x+3\,{\dot K_c\over K_c}\,\dot x+
\mu^2\,x={1\over{K_c}^6\,x^3}
\ ,
\label{pinney}
\ee
with suitable initial conditions.
\par
The system admits an invariant ground state (vacuum) 
$\ket{0,\tau}_b$ defined by
\be
\hat I_b(\tau)\,\ket{0,\tau}_b=0 
\ ,
\label{vacuum}
\ee
and one can define an invariant basis of states 
${\cal B}=\{\ket{n,\tau}_b\}$ with a Fock space
\begin{eqnarray}
\ket{n,\tau}_{bs}&\equiv& 
{(\hat I_b^\dagger)^n\over\sqrt{n!}}\,\ket{0,\tau}_{bs} 
\nonumber \\
&=&e^{-i\,n\,\Theta}\,
{(\bc)^n\over\sqrt{n!}}\,\ket{0,\tau}_{bs} 
\nonumber \\
&=&e^{i\,(\varphi_0-n\,\Theta)}\,
{(\bc)^n\over\sqrt{n!}}\,\ket{0,\tau}_b 
=
e^{i\,\varphi_n(\tau)} \,\ket{n,\tau}_b
\ ,
\end{eqnarray}
where $\varphi_0$ may be replaced by $-\Theta/2$ and
\be
\begin{array}{c}
\b\,\ket{n,\tau}_b=\sqrt{n}\,\ket{n-1,\tau}_b
\\
\\
\bc\,\ket{n,\tau}_b=\sqrt{n+1}\,\ket{n+1,\tau}_b
\ .
\end{array}
\ee
Thus, since $[\b,\bc]=1$, in the following we will refer to $\b$ 
and its Hermitian conjugate $\bc$ as the {\em invariant annihilation} 
and {\em creation} operators, and introduce the hermitian quadratic 
invariant
\be
\hat I_c \equiv\hbar\,\left(\hat I_b^\dagger\,\hat I_b+\frac{1}{2}\right) 
=\hbar\,\left(\hat b^\dagger \,\hat b + \frac{1}{2}\,\right)
\ ,
\ee
with $\hat b^\dagger\,\hat b$ the {\em invariant number} operator.
\par
The {\sl particle annihilation} and {\sl creation} operators $\a$ and 
$\a^\dagger$ are defined by
\be
\a(\tau)\equiv \sqrt{\mu\,{K_c}^3\over2\hbar}\,
\left[\hat\phi+i\,{\hat\pi_\phi\over\mu\,{K_c}^3}\right]
\ ,
\ee
and the Hamiltonian operator is then
\be
\hat H_M(\tau)=\hbar \mu \left[\hat N+{1\over 2}\right]
\ ,
\ee
where $\hat N\equiv \ac\,\a$ is the {\sl particle number}
operator which counts the number of quanta of the scalar field
$\phi$. 
One also has a corresponding vacuum $\ket{0,\tau}_a$ defined by 
$\hat a \, \ket{0,\tau}_a = 0$ and a complete set of eigenstates 
${\cal A}=\{\ket{n,\tau}_a\}$,
\be
\ket{n,\tau}_a\equiv {(\ac)^n\over\sqrt{n!}}\,\ket{0,\tau}_a
\ ,
\ee
such that
\be
\begin{array}{c}
\a\,\ket{n,\tau}_a=\sqrt{n}\,\ket{n-1,\tau}_a
 \\
\\
\ac\,\ket{n,\tau}_a=\sqrt{n+1}\,\ket{n+1,\tau}_a
\ ,
\end{array}
\ee
and $[\a,\ac]=1$.
\par
The two different Fock space basis are related. Indeed one has
\be
\begin{array}{c}
\hat\phi=\sqrt{\hbar\over 2}\,x\,(\b+\bc) \\
 \\
\hat\pi_\phi=\sqrt{\hbar\over 2}\,\left[{i\over x}\,(\bc-\b)
+\dot x\,{K_c}^3\,(\b+\bc)\right]
\ ,
\end{array}
\ee
from which
\be
\left\{\begin{array}{l}
\a=B^\ast\,\b+A^\ast\,\bc \\
 \\
\ac=B\,\bc+A\,\b
\ ,
\end{array}\right.
\label{bogolubov}
\ee
where
\be
\begin{array}{l}
A(\tau)={1\over 2}\,\sqrt{\mu\,{K_c}^3}\,\strut\displaystyle
\left(x-{1\over\mu\,{K_c}^3\,x}-i\,{\dot x\over\mu}\right)   \\
 \\
B(\tau)={1\over 2}\,\sqrt{\mu\,{K_c}^3}\,\strut\displaystyle
\left(x+{1\over\mu\,{K_c}^3\,x}-i\,{\dot x\over\mu}\right)
\ ,
\end{array}
\label{bogo_c}
\ee
are the Bogoliubov coefficients. 
From the above one has an inverse relation,
\be
\left\{\begin{array}{l}
\b=B\,\a - A^\ast\,\ac \\
 \\
\bc=B^\ast\,\ac - A\,\a
\ .
\label{bogo}
\end{array}\right.
\ee
\par
The two basis $\cal A$ and $\cal B$ will coincide in the adiabatic 
limit for which the derivatives of $x\,K_c^{3/2}$ are small. 
Let us suppose that for $\tau=\tau_{_0}$ the two sets
$\{\ket{n,\tau_{_0}}_a\}$ and $\{\ket{n,\tau_{_0}}_b\}$
coincide.
This is achieved if
\be
\b(\tau_{_0})=\a(\tau_{_0}) \ \ \ \Rightarrow\ \ \
\mu \hat I_c(\tau_{_0})=\hat H_M(\tau_{_0})
\ ,
\label{init0}
\ee
which leads to the following initial conditions for the
function $x$
\be
\left\{\begin{array}{l}
x(\tau_{_0})=\left[\mu\,{K_c}^3(\tau_{_0})\right]^{-1/2} \\
 \\
\dot x(\tau_{_0})=0
\ .
\end{array}\right.
\label{init}
\ee
In particular, the second condition in Eq.~(\ref{init}) guarantees
that the state of the system satisfies adiabaticity in the limit
$\tau\to\tau_{_0}$.
\par
In order to study deviations from adiabaticity it is convenient to  
introduce $\sigma = (\mu K_c^3)^{1/2}\,x$ whereupon 
Eq.~(\ref{pinney}) becomes
\be
\frac{1}{\mu^2}\,\ddot\sigma+\left(1-\frac{3\,\ddot K_c}{\mu^2\,K_c}- 
\frac{3\,\dot K_c^2}{4\,\mu^2\,K_c^2}\right)\,
\sigma\equiv\frac{1}{\mu^2}\,\ddot\sigma+\Omega^2\,\sigma
=\frac{1}{\sigma^3}
\ .
\label{pinney2}
\ee
As a first approximation \cite{cv}, which we shall return to when we  
consider the effect of back-reaction, we set $K_c\simeq K_{cl}$ 
which amounts to taking $\expec{\hat H_M}$ as constant and equal 
to $K_0/\kappa$.      
This may be assumed to occur for the time $\tau_0$ for which the adiabatic 
approximation is valid ($\hat a$ and $\hat b$ coincide). 
One then obtains
\be
\Omega=\left[1+\frac{3\,\epsilon\,\delta^2}{4\,(\partial_\eta h_\epsilon)^2}
\right]^{1/2} 
\simeq\left[1+\frac{3\,\epsilon\,\delta^2}{8\,(\partial_\eta h_\epsilon)^2}
\right] 
\ ,
\ee
where $\delta = (\mu\,K_0)^{-1}$. 
In the above, $\delta$ plays the role of an {\sl adiabaticity} parameter
which connects the departure from adiabaticity to the time as the
collapse proceeds ($\tau<\tau_0$).
Indeed the collapse is adiabatic for $\delta \rightarrow 0$, 
in agreement with Eq.~(\ref{cond}).
For the case $\epsilon=0$, one has $\Omega=1$ and the {\em exact}
solutions of the non-linear Eq.~(\ref{pinney2}) 
(with $K_c=K_{cl}$) can be obtained from the expressions displayed below 
by setting $\epsilon=0$.
For a general value of $\epsilon$ ($=0,\pm 1$) the solutions to 
Eq.~(\ref{pinney2}) can be expanded in positive powers of 
$\delta^2$ \cite{lewis},
\be
\sigma = \sum\limits_{n=0}^{\infty}\,\delta^{2n}\,\sigma_n 
\ ,
\ee
where $\sigma_n$ may be expressed in terms of $\Omega$ and its derivatives 
with respect to $\tau$.
To the lowest order in $\delta^2$ one has
\be
\sigma = \Omega^{-1/2} + {\cal O}(\delta^4)\simeq
1-{3\,\epsilon\over 16\,\mu^2\,{K_{cl}}^2} 
\ .
\ee
Further, given a particular solution $\tilde \sigma$ for $\sigma$ (and 
correspondingly for $x$), the general solution to the classical 
equation for $\phi$ obtained from Eq.~(\ref{s1}),
\be
\ddot \phi_c+3\,{\dot K_c\over K_c}\,\dot \phi_c
+\mu^2\,\phi_c=0
\ ,
\ee
is \cite{cole}
\be
\phi_c ={\tilde\sigma\over\sqrt{\mu}}\,K_c^{-\frac{3}{2}}\,
(D\,\cos\Theta + E\,\sin\Theta) 
\ ,
\label{sincos}
\ee
where $D$ and $E$ are constants and $\Theta$ is given in Eq.~(\ref{theta}).
We observe that this solution, obtained here through the adiabatic expansion 
of Ref.~\cite{lewis}, actually coincides with the WKB-type solution of chapter
3.5 of Ref.~\cite{birrell}.
In fact, the non-linear equation (3.103) in \cite{birrell} can be recovered 
from our Eq.~(\ref{pinney2}) by defining $W=1/\sigma^2$ and the vacuum 
corresponding to $\sigma$ as given above is the 
{\em second adiabatic order vacuum} in the terminology of Ref.~\cite{birrell}.
\par
The introduction of the eigenstates of adiabatic invariants allows one to 
consider the effect of particle production due to the variation of the 
metric on the evolution of matter.
In order to see this, it is convenient to introduce coherent states in 
the $\hat b$ modes,
\be
\ket{\alpha,\tau}_{bs}=e^{-|\alpha|^2/2}\,
\sum\limits_{n=0}^\infty\,{\alpha^n\over\sqrt{n!}}\,
\ket{n,\tau}_{bs}=
e^{-|\alpha|^2/2 - i\Theta/2}\,
\sum\limits_{n=0}^\infty\,{\alpha^n\,e^{-i\,n\,\Theta}\over\sqrt{n!}}\,
\ket{n,\tau}_b 
\ ,
\label{coherent}
\ee
where $\alpha=u+i\,v$ is an arbitrary constant.
We note that the modulus squared of the coefficient of $\ket{n,\tau}_b$ 
in the above equation satisfies a Poisson distribution with a maximum 
at $n=|\alpha|^2$, average value $|\alpha|^2$ and standard deviation 
$|\alpha|$. 
If for some time $\tau_o$ the adiabatic approximation is valid then 
the above Eq.~(\ref{coherent}) will also be a coherent state in the 
${\cal A}$ basis.
However, subsequently it will correspond to a {\sl squeezed} state 
in the ${\cal A}$ basis, since the ${\cal B}$ basis is related to the 
${\cal A}$ basis through a Bogolubov transformation Eq.~(\ref{bogo}).
\par
One may evaluate the expectation value of the diverse quantities of 
physical interest with respect to the above state. 
On defining $\expec{\hat O}_{bs} \equiv \bra{\alpha,\tau}\,\hat O \,
\ket{\alpha,\tau}_{bs}$ one obtains
\be
\phi_c \equiv \expec{\hat \phi}_{bs} = 
\sqrt{2\,\hbar}\,|\alpha|\,x\,\cos(\Theta-\beta)
\ ,\ \ \ \ \ \ \ \tan\,\beta\,=v/u
\ ,
\label{ampl}
\ee
in agreement with Eq.~(\ref{sincos}).
Similarly for the momentum $\hat \pi_\phi$ one obtains
\be
\pi_{\phi,c} \equiv
\expec{\hat\pi_\phi}_{bs} = 
\sqrt{2\,\hbar}\,|\alpha|\,\left[
{K_c}^3\,\dot x\,\cos(\Theta-\beta)
-{1\over x}\,\sin(\Theta-\beta)\right]
\ .
\ee
Further,
\be
\begin{array}{c}
\expec{(\Delta\hat\phi)^2 }_{bs}\equiv
\expec{(\hat \phi-\phi_c)^2}_{bs}
=\strut\displaystyle{\hbar\over 2}\,x^2  \\
 \\
\expec{(\Delta\hat\pi_\phi)^2}_{bs}\equiv
\expec{(\hat \pi_\phi-\pi_{\phi,c})^2}_{bs}
=\strut\displaystyle{\hbar\over 2}\,
\strut\displaystyle\left({1\over x^2}+{K_c}^6\,\dot x^2\right)
\ ,
\end{array}
\ee 
both of which are independent of $\alpha$ and lead to
\be
\expec{(\Delta\hat\phi)^2 }_{bs}^{1/2}
\,\expec{(\Delta\hat\pi_\phi)^2}_{bs}^{1/2} = 
{\hbar\over 2}\,\sqrt{1+{K_c}^6\,x^2\,\dot x^2} 
\ .
\ee
Again this result is independent of $\alpha$ and we note that 
the uncertainty relation above is a minimum in the adiabatic 
approximation when the two Fock spaces ${\cal A}$ and ${\cal B}$ coincide.
One also has
\be
\expec{\hat I_c }_{bs} = \hbar \left( |\alpha|^2 + \frac{1}{2} \right) 
\ee 
\begin{eqnarray}
\expec{\hat H_M(\tau)}_{bs} &=& \frac{1}{2\,K_c^3}\,
\left[\expec{\hat \pi_\phi ^2}_{bs}+
\mu^2\,K_c^6\expec{\hat \phi^2}_{bs} \right] \nonumber \\
&=& \frac{1}{2\,K_c^3}\,\left[ \frac{\hbar}{2}
\left( \dot x ^2 \, K_c^6 + \frac{1}{x^2} + \mu^2 \, x^2 \, K_c^6\right)
+\expec{\hat \pi_\phi}_{bs}^2+
\mu^2 K_c^6 \, \expec{\hat \phi}_{bs}^2 \right]
\ ,
\end{eqnarray}
and, of course, in the adiabatic approximation,
\be
\mu\,\expec{\hat I_c }_{bs} \simeq \expec{\hat H_M(\tau) }_{bs}
\ .
\ee 
\par
It is of a particular interest to examine the behaviour of the above 
quantities when one has (small) deviations from the adiabatic 
approximation.
This may be achieved by considering corrections of ${\cal O}(\delta^2)$ 
to the adiabatic approximation and will allow us to see the effect of 
matter not following gravitation adiabatically and the corresponding 
back-reaction on gravitation.
In particular one has
\be
\phi_c \simeq |\alpha| \left[ \frac{2\,\hbar}{\mu\,(K_0\,\partial_\eta\,
h_\epsilon)^3} \right]^{1/2}\,\left[1-
\frac{3\,\epsilon \, \delta^2}{16\,\mu^2\,(\partial_\eta\,h_\epsilon)^2}
\right]\,\cos(\Theta-\beta) 
\ ,
\ee
where now
\be
\Theta \simeq \frac{1}{\delta}\, \int_{\eta_0(\tau_0)}^\eta d\zeta \, 
\partial_\zeta \,h_\epsilon \, \left[ 1 + \frac{3\,\epsilon\,\delta^2}
{4\,(\partial_{\zeta}\,h_\epsilon)^2} \right]^{1/2}  
\ .
\ee
Further,
\begin{eqnarray}
\pi_{\phi,c} &\simeq& 
-\sqrt{\frac{2\,\hbar}{\delta}}\,|\alpha|\,K_0\,
(\partial_\eta\,h_\epsilon)^{\frac{3}{2}}\, \left\{ 
\left[ 1 + \frac{3\,\epsilon\,\delta^2}{16\,
(\partial_\eta\,h_\epsilon)^2} \right]
\sin(\Theta-\beta) \right. \nonumber \\
&&
\phantom{-\sqrt{\frac{2\,\hbar}{\delta}}\,|\alpha|\,K_0\,
(\partial_\eta\,h_\epsilon)^{\frac{3}{2}}\,\{}
+\left.
\frac{3}{2}\,\frac{\delta\,\partial_\eta^2\,h_\epsilon}{(\partial_\eta\,
h_\epsilon)^2} \,
\cos(\Theta-\beta) \right\} 
\ ,
\end{eqnarray}
and
\be
\expec{(\Delta\hat\phi)^2 }_{bs} \simeq
\frac{\hbar\,\delta}{2\,K_0^2\,(\partial_\eta\,
h_\epsilon)^3} \, \left[ 1 - \frac{3\,\epsilon\,\delta}{8\,
(\partial_\eta\,h_\epsilon)^2} \right]
\ee
\be
\expec{(\Delta\hat\pi_\phi)^2}_{bs} \simeq 
\frac{\hbar}{2\delta}\,K_0^2\, (\partial_\eta\,h_\epsilon)^3 \left\{ 
1 + \frac{3\,\delta^2}{4\,
(\partial_\eta\,h_\epsilon)^2} \left[ \frac{\epsilon}{2} + 3\,
\frac{(\partial_\eta^2\,h_\epsilon)^2}{(\partial_\eta\,h_\epsilon)^2} \right]
\right\}
\ee
\be
\expec{(\Delta\hat\phi)^2 }_{bs}^{1/2}
\,\expec{(\Delta\hat\pi_\phi)^2}_{bs}^{1/2} \simeq 
\frac{\hbar}{2} \, \left[ 1 + \frac{9}{4} \delta^2 \frac{(\partial_\eta^2
\,h_\epsilon)^2}{(\partial_\eta\,h_\epsilon)^4} \right]^{1/2}
\ .
\label{ind0}
\ee
Also, the expectation value of the matter Hamiltonian is given by
\begin{eqnarray}
\expec{\hat H_M(\tau)}_{bs} &\simeq& \frac{\hbar \mu}{2} + 
\hbar \mu |\alpha|^2 
\left\{ \left[ \sin(\Theta - \beta) + \frac{3}{2}\delta 
\frac {\partial_\eta^2\,h_\epsilon}{(\partial_\eta\,h_\epsilon)^2} 
\cos(\Theta - \beta) \right]^2 + 
\cos^2(\Theta - \beta) \right\}
\nonumber \\
&=& \hbar \mu \left( |\alpha|^2 + \frac{1}{2} \right) 
+ 3\, \hbar \mu \,|\alpha|^2 
\delta \frac {\partial_\eta^2
\,h_\epsilon}{(\partial_\eta\,h_\epsilon)^2}\, \sin(\Theta - \beta) \,
\cos(\Theta - \beta) \nonumber \\
&\equiv& \frac{K_0}{\kappa} + \Delta \expec{\hat H_M}_{bs}
\ ,
\label{pp0}
\end{eqnarray}
where only the first term $K_0/\kappa$ on the r.h.s. of Eq.~(\ref{pp0}) 
survives in the adiabatic approximation and the second term is associated 
with particle production.
\par
In order to better understand our results, in particular Eq.~(\ref{ind0}) 
and Eq.~(\ref{pp0}), we may consider $\eta^3$ small and $\gg\delta$
corresponding to $\mu^{-1}\ll\tau\ll \tau_0$, which implies that we are almost 
adiabatic (we only consider small deviations from adiabaticity).
It is then straightforward to see that, in such a case, from 
Eq.~(\ref{ind0}),
\be
\expec{(\Delta\hat\phi)^2 }_{bs}^{1/2}
\,\expec{(\Delta\hat\pi_\phi)^2}_{bs}^{1/2} \simeq 
\frac{\hbar}{2} \left[1 + {\cal O}\left(\frac{\delta^2}{\eta^6}\right)
\right] 
\ ,
\label{ind}
\ee
which of course shows that as $\eta$ becomes smaller (during the collapse), 
matter becomes less and less classical.
On the other hand, from Eq.~(\ref{pp0}) one obtains
\be
\Delta\expec{\hat H_M}_{bs} \simeq \hbar\,\mu \,|\alpha|^2 \, {\cal O} 
\left(\frac{\delta}{\eta^3}\right)
\ ,
\label{pp}
\ee
again showing that, as $\eta$ becomes small, one has an increasing 
production of matter.
Clearly such a production will induce a back-reaction on gravitation 
and $K_c$ will change \cite{cita1} from $K_{cl}$ to $K_{cl}+\Delta K_c$, 
where $\Delta K_c$ is $K_0\,{\cal O}(\delta/\eta^3)$. 
Since we are considering an exterior Schwarzschild metric and the time at 
which the horizon is crossed by the last {\sl shell} of matter is given 
by $2\, M = \rho_0\,K_c(\eta_H)$, the presence of the additional term 
$\Delta K_c$ will lead to $\rho_0\,K_{cl}(\eta_H) \simeq 2\,M - \rho_0\,K_0\,
{\cal O}(\delta/\eta_H^3)$.
This of course implies that the horizon is crossed at a later time 
(smaller value of $\eta$) and that the back-reaction (matter production)
slows down the collapse.
The effect of particle creation on a Friedmann-like collapse and the
possible avoidance of the cosmological singularity due to quantum
effects have been studied in Ref.~\cite{parker}.
\par
In obtaining the above results we have neglected the r.h.s. of 
Eqs.~(\ref{wdw_g})-(\ref{wdw_m}) which are associated with fluctuations 
in the particle production due to the variation of the metric 
and the term ${\cal O}(\hbar)$ in our introduction of time 
in Eq.~(\ref{eta}). 
Let us estimate these effects and compare them with particle production. 
For this purpose, order of magnitude estimates of the r.h.s. are 
sufficient. 
They can be obtained by just evaluating the r.h.s. for a state 
$\ket{n=N,\tau}_b$ such that $N=|\alpha|^2$, since it is for this 
value that the (modulus) coefficients of the expansion in 
Eq.~(\ref{coherent}) are peaked and corresponds to the amplitude of 
the classical oscillations (see Eq.~(\ref{ampl})). 
For the r.h.s. of Eq.~(\ref{wdw_g}) one obtains
\begin{eqnarray}
\kappa \hbar^2 \expec{N|\stackrel{\leftarrow}{\frac{\partial}{\partial K}} 
( 1 - \ket{N}\bra{N}
) \frac{\partial}{\partial K} | N} &=& 
\frac{\kappa}{{\dot K}^2} \sum_{L \ne N} \,
\expec{N|\hat H_M|L}\expec{L|\hat H_M|N} \nonumber \\
&\simeq& 
K_{cl}\,\Delta \expec{\hat H_M}_{bs} {\cal O}\left(\frac{\delta}{\eta^3}
\right) 
\ ,
\label{val_g}
\end{eqnarray}
whereas for the modulus of the r.h.s. of Eq.~(\ref{wdw_m}) one has
\begin{eqnarray}
&&\left| - \frac{\hbar^2 \kappa}{2} \left[ \left(
\frac{\partial^2}{\partial K^2} - \langle N|\frac{\partial^2}{\partial K^2}
|N\rangle \right) - 2 \langle N|\frac{\partial}{\partial K} |N\rangle
\left(
\frac{\partial}{\partial K} - \langle N|\frac{\partial}{\partial K}
|N \rangle \right) \right] |N \rangle \right| 
\nonumber \\
&&=\frac{\hbar^2 \kappa}{2} \left| \sum_{L \ne N} \frac{\ket{L}}{ \hbar 
(N - L)} \,\left[ \bra{L}\frac{\partial^2 \hat I_c}{\partial K^2}\ket{N} 
+\sum_{P \ne N,L} \hbar (L-P)\expec{L|\hat H_M|P}\expec{P|\hat H_M|N}\right]
\right| \nonumber \\
&&\simeq K_{cl}\,\Delta \expec{\hat H_M}_{bs} {\cal O}
\left(\frac{\delta}{\eta^3} \right)
\ .
\label{val_m}
\end{eqnarray}   
\par 
Further, the corrections of ${\cal O}(\hbar)$ to Eq.~(\ref{eta}) lead 
to an additional term for the Hamilton-Jacobi equation associated 
with Eq.~(\ref{wdw_g}),
\be
\frac{\kappa\,\hbar^2}{2}\,\left[ \frac{3}{4}\,
\frac{{\dot\pi}_K^2}{{\dot K}^2\,\pi_K^2} 
-\frac{\ddot \pi_K}{2\,{\dot K}^2\pi_K} 
+ \frac{{\ddot K}\,{\dot \pi}_K}{2\,\dot K^3\,\pi_K} \right] \simeq 
\frac{1}{|\alpha|^2}\, K_{cl}\, \Delta\expec{\hat H_M}_{bs} 
{\cal O}\left(\frac{\delta}{\eta^3} \right) 
\ ,
\label{val_pre}
\ee
whereas the modulus of the corresponding contribution to the r.h.s. of 
Eq.~(\ref{wdw_m}) is (omitting $\tilde \psi$)
\be
\kappa\,\hbar^2\,\left|\frac{\dot \pi_K}{2\,\dot K\,\pi_K}\,\left( 
\frac{\partial}{\partial K} - \bra{N}\frac{\partial}{\partial K}
\ket{N}\right)\,\ket{N}\right| \simeq 
K_{cl}\,\Delta\expec{\hat H_M}_{bs} 
{\cal O}\left(\frac{\delta}{\eta^3} \right) 
\ ,
\ee
which is again of higher order with respect to 
$K_{cl} \Delta\expec{\hat H_M}_{bs}$. 
\par
The above results Eqs.~(\ref{val_g})-(\ref{val_m}) are associated with 
particle production due to the metric variations. 
In particular we see from Eq.~(\ref{ind}) that, as the collapse proceeds 
the evolution of matter becomes less and less classical and one has 
the production of particles (Eq.~(\ref{pp})) which, as a consequence, 
slows down the collapse (back-reaction).
Further one has lower order contributions, Eqs.~(\ref{val_g})-(\ref{val_m}), 
which are associated with fluctuations in particle productions and 
affect both the evolution of gravitation and matter and corrections 
to the classical gravitational motion (see Eq.~(\ref{val_pre})).   
\section{Conclusions}
\label{concl}
The matter-gravity system lends itself to a study analogous to that 
employed in molecular-dynamics where one also has two mass (or time) 
scales. 
We previously applied this analogy to the collapse of a sphere of dust 
in the adiabatic approximation, with the matter equation of motion 
being solved while {\sl freezing} the gravitational degrees of 
freedom.
In this note we have generalized such a treatment, to allow for the 
time dependence induced in the matter Hamiltonian by the time variation 
of the gravitational (metric) degree of freedom. 
\par
Time has been introduced by considering the semiclassical (WKB) limit for 
gravitation while the matter equation of motion was solved by neglecting 
fluctuations and employing the method of adiabatic invariants. 
Such invariants, in contrast with the Hamiltonian, have time-independent 
eigenvalues. 
In particular, in our case, which corresponds to a harmonic oscillator 
with time dependent mass, on {\sl freezing} the gravitational degree 
of freedom one reproduces the adiabatic results \cite{cv}.
The adiabatic invariants are then immediately related to the usual 
harmonic oscillator (particle) creation and destruction operators 
and its Hamiltonian.
\par
In general the (non-hermitian) adiabatic invariants are related to 
the usual creation and destruction operators through a Bogolubov 
transformation and therefore do not correspond to the latter.
In analogy with the particle destruction and creation operators, 
however, one can construct coherent states of the adiabatic invariants.
We have also evaluated the expectation values of diverse quantities 
of physical interest with respect to such states and shown, for example, 
that the expectation value of the scalar field operator is a solution 
to the classical equation of motion.
\par
In order to obtain some estimate of the corrections to the adiabatic 
approximation we have determined the diverse quantities to lowest order 
in $\delta$ (the ratio of the Compton wavelength of the scalar particle 
to the Schwarzschild radius of the black hole) and for small enough times 
$\eta$ (nonetheless satisfying $\eta^3 > \delta$). 
This has shown that the dominant term (${\cal O}(\delta/\eta^3)$) comes 
from particle creation in the average matter Hamiltonian, whereas the 
corrections due to the fluctuations in the number of produced particles, 
in the uncertainty principle or in the semiclassical approximation are 
of higher order (${\cal O}(\delta^2/\eta^6)$). 
To this order of approximation (see Eq.~(\ref{pp})) there is no distinction 
between the different three geometries ($\epsilon=0, \pm 1$).
\par
Clearly the production of matter will influence the collapse and one may 
consider the associated back-reaction. 
As we have mentioned, the leading term is of order 
${\cal O}(\delta/\eta^3)$.
It comes from the expectation value of the matter Hamiltonian with 
respect to the adiabatic invariant coherent states and we have seen 
that it leads to a slowing down of the collapse, as expected. 
It would be of interest to also consider the effect of matter being 
radiated away, which would need the extension of our results to a Vaidya, 
rather than a Schwarzschild metric.

\begin{thebibliography}{99}
%
\bibitem{lund}
Lund F 1973 {\em Phys. Rev.} D {\bf 8} 3253
%
\bibitem{cv}
Casadio R and Venturi G 1996 {\em Class. Quantum Grav.} {\bf 13} 2715
%
\bibitem{oppenheimer}
Oppenheimer J R and Snyder H 1939 {\em Phys. Rev.} {\bf 56} 455
%
\bibitem{brout}
Brout R 1987 {\em Found. Phys.} {\bf 17} 603
%
\bibitem{bv}
Brout R  and Venturi G  1989 {\em Phys. Rev.} D {\bf 15} 2436
%
\bibitem{kiefer}
Kiefer C 1994 {\em Canonical Gravity -- From Classical to Quantum}
ed Ehlers J and Friedrich H (Berlin: Springer)
%
\bibitem{kim}
Kim S P 1995 {\em Phys. Lett.} {\bf 104A} 359 and
{\em Phys. Rev.} D {\bf 52} 3382
%
\bibitem{datta}
Datta D P 1993 {\em Phys. Rev.} D {\bf 40} 574 and
{\em Gen. Rel. Grav.} {\bf 27} 341
%
\bibitem{bertoni}
Bertoni C Finelli F and Venturi G 1996 {\em Class. Quantum Grav.} 
{\bf 13} 2375
%
\bibitem{stephani}
Stephani H 1990 {\em General Relativity} (Cambridge: Cambridge Univ. Press)
%
\bibitem{hk}
Hajicek P and Kijowski J 1998 {\em Phys. Rev.} D {\bf 57} 914
%
\bibitem{os}
Casadio R 1998 {\em Hamiltonian formalism for the Oppenheimer-Snyder
model}, preprint gr-qc/9804021
%
\bibitem{madsen}
Madsen M S 1988 {\em Class. Quantum Grav.} {\bf 5} 627 
%
\bibitem{parker}
Parker L and Fulling S A 1973 {\em Phys. Rev.} D {\bf 7} 2357
%
\bibitem{moss}
Goncalves S and Moss I G 1997 {\em Class. Quantum Grav.} {\bf 14} 2607
%
\bibitem{lewis-ries}
Lewis H R and Riesenfeld W B 1969 {\em J. Math. Phys.} {\bf 10} 1458
%
\bibitem{gao}
Gao X-C Xu J-B and Qian T-Z 1991 {\em Phys. Rev.} A {\bf 44} 7016
%
\bibitem{n_cv}
In Ref.~\cite{cv} the true action is shown to be $V_s\,S$,
where $S$ is given in Eq.~(\ref{s1}) and 
$V_s(\rho_0)=4\,\pi\,\int_0^{\rho_0}\rho^2\,d\rho/
\sqrt{1-\epsilon\,\rho}$ is the coordinate volume of the sphere.
However, to simplify the notation in the present paper we have set 
$V_s$ to one, since it can be factored out in the Wheeler-DeWitt
equation by making use of the canonical transformation
$\pi_{_K}\to\pi_{_K}/{V_s}^{1/2}$, $K\to K\,{V_s}^{1/2}$,
$\pi_\phi\to\pi_\phi\,{V_s}^{1/2}$, $\phi\to\phi/{V_s}^{1/2}$,
and thus it plays no role in the dynamics.
%
\bibitem{banks}
Banks T 1985 {\em Nucl. Phys} B {\bf 249} 332
%
\bibitem{cita}
In Ref.~\cite{cv} we gave an explicit expression for the wave 
function $\tilde\psi$ which is peaked at $K_{cl}$ given
in Eq.~(\ref{K_cl}).
However, in the present context we let $K_c=K_c(\tau)$ be a solution 
to the Hamilton-Jacobi equation for a given $S_{eff}$, Eq.~(\ref{s_eff}),
where we allow for a time dependence of $\expec{\hat H_M}$ in contrast with 
$K_0=\,$constant in $S_{cl}$, Eq.~(\ref{s_cl}). 
In the adiabatic approximation $K_c$ and $K_{cl}$ coincide.
%
\bibitem{lewis}
Lewis H R 1968 {\em J. Math. Phys.} {\bf 9} 1976
%
\bibitem{cole}
Prince G E and Eliezier L J 1980 {\em J. Phys} {\bf A 13} 815;
Lutzky M 1978 {\em Phys. Lett.} {\bf 68 A} 3
%
\bibitem{birrell}
Birrell N D and Davies P C W 1982 {\em Quantum Fields in
Curved Space} (Cambridge Univ. Press)
%
\bibitem{cita1}
See \cite{cita}
%
\end{thebibliography}
\end{document}